\documentstyle[aps,pre,multicol,twoside,epsf]{revtex}
\makeatletter

  \def\refrule{%
  \end{multicols}\widetext\vglue8pt %
  \hskip10.25pc\rule{20pc}{.1mm}\hfill %
  \vglue.14cm\begin{multicols}{2}\narrowtext %
  }
  \def\references{%
  \list{\@biblabel{\arabic{enumiv}}}%
  {\labelwidth\WidestRefLabelThusFar  \labelsep2pt %
  \leftmargin\labelwidth %
  \advance\leftmargin\labelsep %
  \ifdim\baselinestretch pt>1 pt %
  \parsep  4pt\relax %
  \else   %
  \parsep  0pt\relax %
  \fi
  \itemsep\parsep %
  \usecounter{enumiv}%
  \let\p@enumiv\@empty
  \def\theenumiv{\arabic{enumiv}}%
  }%
  \let\newblock\relax %
  \sloppy\clubpenalty4000\widowpenalty4000
  \sfcode`\.=1000\relax
  \ifpreprintsty\else\small\fi
  }
  \newfont{\rmtop}{cmr10 at 9pt}
 
  \def\AUTHORS{\rmtop FERENC IGL\'OI AND  LO\"{I}C TURBAN}
  \def\JOURNAL{PHYSICAL REVIEW E}
  \def\DATE{2002}
  \def\TITLE{\rmtop FIRST- AND SECOND-ORDER PHASE TRANSITIONS ...}
  \def\VOLUME{{\small\bf 66}}
  \def\NUMBER{036140}
  \def\PAGE{1}

  \def\ps@plain{%
  \gdef\@oddhead{\ifnum\thepage=\PAGE%
  {\hfill\rmtop\JOURNAL\ \VOLUME,\ \NUMBER\ (\DATE)\hfill}%
  \else{\rmtop\TITLE\hfill \JOURNAL\ \VOLUME,\ \NUMBER\ (\DATE)}\fi}%
  \gdef\@evenhead{\rmtop\AUTHORS\hfill \JOURNAL\ \VOLUME,\ \NUMBER\ (\DATE)}%
  \gdef\@oddfoot{\ifnum\thepage=\PAGE%
  {\rmtop\hfill\VOLUME\ \NUMBER-\thepage\hfill}
  \else{\rmtop\hfill\NUMBER-\thepage\hfill}\fi}%
  \gdef\@evenfoot{\rmtop\hfill\NUMBER-\thepage\hfill}%
  }%
\makeatother
\begin{document}

\title{First- and second-order phase transitions in scale-free networks}

\author{Ferenc Igl\'oi$^{1,2,3}$ and Lo\"{\i}c Turban$^{3}$}

\address{
$^1$ Research Institute for Solid State Physics and Optics, 
H-1525 Budapest, P.O. Box 49, Hungary\\
$^2$ Institute of Theoretical Physics,
Szeged University, H-6720 Szeged, Hungary\\
$^3$Laboratoire de Physique des Mat\'eriaux, Universit\'e Henri Poincar\'e (Nancy 1),
F-54506 Vand\oe uvre l\`es Nancy, France
}

\date{Received 26 June 2002}

\maketitle

\begin{abstract}
We study first- and second-order phase transitions of ferromagnetic lattice
models on scale-free networks, with a degree exponent $\gamma$.
Using the example of the $q$-state Potts model we derive a general self-consistency
relation within the frame of the Weiss molecular-field approximation, which presumably
leads to exact critical singularities. Depending on the value of $\gamma$,
we have found three different regimes of the phase diagram. As a general trend
first-order transitions soften with decreasing $\gamma$ and the critical singularities
at the second-order transitions are $\gamma$ dependent. 
\end{abstract}

\pacs{64.60.Cn, 05.50.+q, 68.35.Rh} 

\newcommand{\bc}{\begin{center}}
\newcommand{\ec}{\end{center}}
\newcommand{\be}{\begin{equation}}
\newcommand{\ee}{\end{equation}}
\newcommand{\beqn}{\begin{eqnarray}}
\newcommand{\eeqn}{\end{eqnarray}}

\begin{multicols}{2}
\narrowtext

Complex networks, which have more complicated connectivity structure than periodic
lattices (PLs) have attracted considerable interest recently\cite{AB01,DM01}. 
This research is motivated
by empirical data collected and analyzed in different fields. Small-world (SW)
networks\cite{WS98}, which can be generated from PLs by replacing a
fraction $p$ of bonds by new random links of arbitrary lengths, are suitable
to model neural networks\cite{neural} and transportation systems\cite{transzp}. 
On the other hand, scale-free (SF)
networks\cite{BA99} are realized among others in social systems\cite{social}, 
in protein interaction networks\cite{protein}, in the
internet\cite{internet} and in the world-wide web\cite{WWW}.
In a SF network the degree distribution $P_D(k)$, where $k$ is the number of links
connected to a vertex, has an asymptotic power-law decay $P_D(k) \sim k^{-\gamma}$,
thus there is no characteristic scale involved. In natural and artificial networks the value
of the degree exponent is usually in the range $2<\gamma<3$\cite{GOH}.

Cooperative processes such as spread of epidemic disease\cite{epidemic},
percolation\cite{percolation}, Ising model\cite{stauffer,ising}, etc.
have also been studied in the SW and the SF networks. For SW networks numerical studies
show\cite{SW_phasetr} that any finite fraction
of new, long-range bonds, $p>0$, brings the transition into the classical,
mean-field (MF) universality class. It is understandable since for systems with
long-range interactions the MF approximation is exact. In the SF networks, where
links between remote sites exists, too, at first thought one could expect also
a traditional MF critical behavior. In specific problems, however, it turned out
that it is only true for losely connected networks, when the degree exponent $\gamma$
is large enough. Otherwise the critical singularities of the transition are
model independent, but nonuniversal; the critical exponents continuously depend on
the value of the degree exponent. In particular, for $2<\gamma \le 3$, when $\langle
k^2 \rangle$ is divergent the systems are in their ordered phase for any value
of the control parameter (temperature, percolation probability, transition rate, etc.),
and the critical properties can be investigated in the limit of infinitely strong
fluctuations.

Till now investigations on cooperative processes in the SF networks are almost exclusively limited
to continuous phase transitions. However, in many problems the phase transitions on
PLs are first order and it seems natural to ask what happens with
these transitions on the SF networks? There is a general tendency that the
discontinuities (e.g., the latent heat) in the pure system are reduced
due to inhomogeneities, which often change the transition into a continuous one.
This has been observed in the vicinity of free surfaces\cite{lipowsky},
when there are missing bonds, or in the bulk when random\cite{cardy} or
aperiodic\cite{aperiodic} perturbations are
present. 

In the present paper, we investigate this issue on the SF networks. In particular
we are interested in the combined effect of strong connectivity and irregularities,
present in the SF networks, on the properties of discontinuous phase transitions.
In the actual calculations we start with the ferromagnetic $q$-state Potts model and solve
it in the frame of the Weiss molecular-field approximation, which represents a lattice
version of the MF method. Then we generalize this procedure for any lattice model and
show how the MF equation on the SF networks can be deduced from the corresponding one
for PLs. The MF equation is analyzed
by standard methods\cite{landau} and the properties of the phase transitions, in particular those
related to a first- to second-order crossover are calculated. Since the MF method is
expectedly exact for the SF networks our results are presumably exact. 

In the following, we consider the $q$-state ferromagnetic Potts model\cite{wu} defined
by the Hamiltonian:
\be
-\frac{H}{k_B T}= \sum_{\langle ij \rangle} K_{ij}\delta(s_i,s_j) +
\sum_i h_i \delta(s_i)
\label{hamilton}
\ee
in terms of Potts spin variables, $s_i=0,1, \dots, q-1$, at site $i$.
The interaction $K_{ij}$ is equal to $K>0$ if the bond $\langle ij \rangle$
is occupied and zero, otherwise. As is well known, the Potts
model contains as special cases the Ising model for $q=2$ and the
bond percolation problem in the limit $q \to 1$.
On regular, $d$-dimensional lattices in the absence of external fields the
phase transition of the homogeneous model is first order, as in the MF theory,
for $q>q_c(d)$ and continuous for $q \le q_c(d)$ where $q_c(2)=4$,
$q_c(3) \lesssim 3$ and $q_c(d \ge 4)=2$.

To find the thermodynamical properties of the model we use the MF method, when the
problem is transformed to a set of independent spins in the presence of effective
local fields, which are created by the nearest neighbors. The partition function
is then given as a product of single site contributions, $Z=\prod_i z_i$,
and the free energy $F$ takes the form,
\beqn
-\frac{F}{k_B T}=\sum_i \sum_j \frac{K_{ij}}{2q}
[1-2 m_j - (q-1)m_i m_j]\cr
 + \sum_i \ln \left[ \exp\big(\sum_j K_{ij} m_j + h_i\big) +q-1 \right]\;.
\label{F}
\eeqn
Here we introduced the local magnetization $0 \le m_i \le 1$ as
\be
m_i=\frac{q\langle \delta(s_i) \rangle -1}{q-1}\;,
\label{m_i}
\ee
the value of which follows from the extremal condition
of the free energy $\partial F/\partial m_i=0$, leading to a set of self-consistency
(SC) equations:
\be
\sum_i K_{ij} m_i= \sum_i K_{ij} \frac{\exp\left(\sum_j K_{ij} m_j +h_i\right)-1}
{\exp\left(\sum_j K_{ij} m_j +h_i\right)+q-1}\;.
\label{self_cons}
\ee

On a PL with coordination number, $z$, $m_i=m_0$ and $h_i=h$
one obtains the relation,
\be
m=G(zKm_0+h),\quad G(x)=G_P(x)=\frac{e^x -1}{e^x +q -1}\;,
\label{m_hom}
\ee
which is compatible with a first-order (second-order) transition for $q>2$ ($q \le 2$).

For a SF network we consider no correlations (anticorrelations) between the degrees
of connected sites and assume that the probability of having a link between sites
$i$ and $j$, $p_{ij}$ is proportional to the number of links connected to each sites,
i.e., $ p_{ij} \sim k_i k_j$. Furthermore, in the spirit of the MF method we replace
the interaction, $K_{ij}$, in Eq.~(\ref{self_cons}) by its average value\cite{note},
$K_{ij}\to K (k_i k_j/\sum_i k_i)$.
Now in terms of the average order parameter, $m=\sum_i k_i m_i/\sum_i k_i$ and
for homogeneous field $h_i=h$ one obtains
from Eq.~(\ref{self_cons}) the SC equation for the SF networks:
\be
m=\! \int \! dk P_D(k) k\, G(kKm+h)/\langle k \rangle=G_{SF}(Km,h)\;,
\label{m_SF}
\ee
where summation with respect to $i$ is replaced by integration over the degree $k$
as $(1/N)\sum_i \to \int dk P_D(k)$, where $N$ is the number of
vertices. Note that the SC equations for PLs
in Eq.~(\ref{m_hom}) and for the SF networks in Eq.~(\ref{m_SF}) are in similar form,
and the SC function for networks, $G_{SF}(Km,h)$ is directly related to that in
a PL, $G(x)$. This latter transformation, as given in Eq.~(\ref{m_SF})
remains the same for any type of lattice model. Therefore, Eq.~(\ref{m_SF})
sets a direct connection between the MF solutions in PLs and in the SF
networks and thus it is a fundamental relation.

Next, we turn to analyze the critical behavior of the SF networks compatible with the general
SC equation in Eq.~(\ref{m_SF}). First, we recall that the
SC function, $G(x)$ is monotonically increasing
from $0$ to $1$ as $x$ varies from $0$ to $\infty$ and the first few terms of its
Taylor expansion, $G(x)=\sum_{n=1} a_n x^n$  are essential for the properties of the
phase transition\cite{landau}. For the Potts model the first three coefficients are given by
$a_1=1/q$, $a_2=(q-2)/(2q^2)$, and $a_3=(q^2-6q+6)/(6q^3)$. For the SF networks, the analogous
SC function, $G_{SF}(Km,h)$, is generally not analytical due to singularities caused by
integration over the degree distribution. For small $m$ (and for small $h$) it has
generally a regular part which is a polynome of finite degree
$\tilde{n}$ where $\tilde{n}$ is the largest natural number smaller than $\gamma-2$:
\be
G^r_{SF}(Km,h)=\sum_{n=1}^{\tilde{n}} a_n \frac{\langle k^{n+1} \rangle}{\langle k \rangle}
(K m)^n+a_1 h\;,
\label{G_small}
\ee
and a singular contribution, which in the small $m$ limit is given by
\be
G_{SF}^s(Km)=\left\{ \begin{array}{ll}
a_s (Km)^{\gamma-2}, & \tilde{n}+2<\gamma<\tilde{n}+3 \cr
B |\ln Km| (Km)^{\tilde{n}+1}, & \gamma=\tilde{n}+3 \cr
\end{array}
\right. 
\label{G_sing}
\ee
Here
\be
a_s = C \int_0^{\infty} \!\!{\rm d} x\, x^{1-\gamma}
 \left[ G(x)-\sum_{n=1}^{\tilde{n}} a_n x^n \right]
\label{a_s}
\ee
and the constants $B$ and $C$ are positive.

Having the small $m$ behavior of the SC function for the SF network at hand,
we can analyze the corresponding critical behavior. Due to the presence of the
singular contribution in Eq.~(\ref{G_sing}) the critical behavior of the
SF network can be different of that in PLs. Generally we can
define three regions of the degree exponent $\gamma$ with different types
of critical behavior. In the following, we are going to describe these
regimes.

$\gamma>\gamma^u$: Conventional mean-field regime. 
If the degree exponent is larger than an upper critical value $\gamma^u$
the critical behavior on the SF network is the same as on a PL.
This happens when the singular term in Eq.~(\ref{G_sing}) does not
modify the usual Landau-type analyzis\cite{landau}. Here, depending on the order
of the transition in the PL, there are two different possibilities:

(i) If the transition in the PL is second order, then
the first two nonvanishing terms of the small $m$ expansion
of $G_{SF}(Km,h)$ should be regular, i.e., $\tilde{n} \ge 2$ and
$\sum_{n=2}^{\tilde{n}} |a_n|>0$. In this case $\gamma^u=n_2+2$, where
$n_2>1$ is the smallest integer for which $a_{n_2}<0$. As an example
for the $(q<2)$-state Potts model (including percolation for which $q=1$)
the upper critical degree exponent is $\gamma^u=4$, since here
$a_2<0$. On the other hand for the Ising model, where $a_2=0$
and $a_3<0$ the upper critical value is $\gamma^u=5$;

(ii) In the second case, when the transition in the PL is
first order, we are looking for the condition that the transition
stays first order in the SF network, too. This will happen,
provided (a) the linear regular term of
$G_{SF}(Km,h)$ exists and (b) the next-order contribution (either regular
or singular) is positive. It is then easy to see that the upper critical
value of $\gamma^u$ is given by the conditions $\tilde{n}=1$ and $a_s(\gamma^u)=0$.
Indeed, for strongly connected networks, when $a_s<0$ the transition is softened
into a second-order one, the unconventional properties of which will be described
later. As an example the first-order transition
of the $(q>2)$-state Potts model in PLs, where $a_2>0$ and $a_3<0$, will
turn into a continuous one on the SF networks for $\gamma<\gamma^u$, where
the upper critical exponent obtained numerically, is
shown in Fig.~1 for different values of $q$.

\begin{figure}[tbh]
\epsfxsize=7truecm
\begin{center}
\mbox{\epsfbox{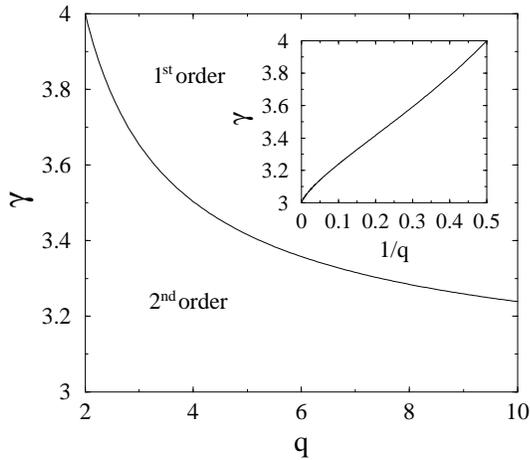}}
\end{center}
\caption{\label{fig1} Regions of first- and second-order phase transitions
for the $q$-state Potts model on scale-free networks with a degree exponent $\gamma$.
In the second-order regime, i.e., below the upper critical value $\gamma^u$ the
singularities are $\gamma$ dependent.
}
\end{figure}

As a general trend $\gamma^u$
is monotonously decreasing with $q$ and approaching the limiting value of
$3$ as $1/q$ for large $q$ (see the inset to Fig. 1). This is consistent with our
expectations; a stronger first-order transition on a PL, which has a larger latent heat,
can be destroyed only in a more connected
network, i.e., with a smaller value of $\gamma$.

Thus we can conclude at this point that for $\gamma \le \gamma^u$ the
effect of the connectivity of the SF network is {\it relevant}, so that the
singularities of the thermodynamical quantities of the system are
different from the conventional mean-field behavior, which can be observed
in PLs, The relevant perturbation region is still divided into two
parts, depending on the position of the singularity: wether it is at
finite or at zero coupling. In the following, we describe these regions.

$3 < \gamma \le \gamma^u$: Unconventional critical regime.
The critical behavior in this regime is due to an interplay between
a regular linear term (which does exists, since $\gamma>3$)
and a negative singular next-to-leading term in the
expansion of $G_{SF}(Km,h)$. As a result the transition is second order
and takes place at a finite coupling, which in the MF method is given by
$K_c={\langle k \rangle}/({\langle k^2 \rangle} a_1)$. Due to the $\gamma$ dependence
of the singular term the singularity of the order parameter is unconventional:
\be
m(K) \sim (\Delta K)^{1/(\gamma-3)},\quad 3 < \gamma < \gamma^u \;,
\label{gamma+}
\ee
where $\Delta K=K-K_c$. At the upper critical value of the degree exponent
$\gamma=\gamma^u$ according to the result in Eq.~(\ref{G_sing}), there are
logarithmic corrections of the form
\be
m(K) \sim \left(\frac{\Delta K}{|\ln \Delta K|} \right)^{1/(\gamma_u-3)},\quad \gamma = \gamma^u \;,
\label{gamma+log}
\ee
at least if the transition in the PL is second order.

If the transition in the PL is first order, then $\gamma=\gamma^u$
corresponds to a tricritical point in the SF network and the tricritical
exponents depend on other details of the degree distribution, such as
the next-to-leading decay exponent.

The behavior of the susceptibility at the transition point is calculated
from the small $h$ expansion of the SC function in Eq.~(\ref{G_small}). Since
the leading contributions are regular, the singularity of the susceptibility
follows the conventional Curie-Weiss law, $\chi(K) \sim 1/|\Delta K|$, and is not
modified by the connectivity effect of the SF network. 

The singularity in the specific heat is directly related to that of the
order parameter and can be deduced from the known relation for
the energy density $e \sim m^2$ valid in the MF theory.

$ \gamma \le 3$: Ordered regime.
If the degree exponent of the SF network is $\gamma \le 3$ (but $\gamma>2$,
in order to ensure a finite average degree, $\langle k \rangle < \infty$),
then the singular properties of the system are exclusively determined
by the leading singular term of the SC function in Eq.~(\ref{G_sing}). As a consequence
the system in the SF network is in its ordered phase at any finite value of the
coupling and singularities take place only at zero coupling (or at infinite
temperature). The order parameter vanishes at $K=0$ as
\be
m(K) \sim K^{(\gamma-2)/(3-\gamma)},\quad 2<\gamma<3\;,
\label{gamma-}
\ee
whereas at the borderline value, $\gamma=3$ there is an essential singularity:
\be
m(K) \sim K^{-1} \exp \left(-1/BK\right), \quad \gamma=3\;.
\label{gamma-log}
\ee
The susceptibility at $K=0$ is generally finite, $\chi=a_1/(3-\gamma)$,
except for $\gamma=3$, when it is divergent as $\chi \sim 1/K$. 

In a finite network with $N$ vertices the order in the system disappears
already at a nonzero coupling, $K_C(N)$, which can be estimated as follows. The
typical value of the largest degree in the finite network, $k_{max}$, is obtained from
the usual condition for extreme events:
$\int_{k_{max}}^{\infty} P_D(k) dk \sim 1/N$, thus
$k_{max} \sim N^{1/(\gamma-1)}$. In a finite system the different moments of $k$ are
also finite, and we obtain for the finite-size scaling behavior of the second moment:
$\langle k^2 \rangle \sim k_{max}^{3-\gamma} \sim N^{(3-\gamma)/(\gamma-1)}$.
From this result the size dependent value of the coupling at the transition point
can be calculated as
\be
K_C(N) \sim \langle k^2 \rangle^{-1} \sim N^{-(3-\gamma)/(\gamma-1)},
\quad 2<\gamma<3 \;,
\label{K*N}
\ee
from which the finite-size scaling behavior of the order parameter at the
transition point
\be
m(N)\sim K_C(N)^{(\gamma-2)/(3-\gamma)} \sim N^{-(\gamma-2)/(\gamma-1)}\;
\ee
follows. For $\gamma=3$ the size dependence of the transition point is logarithmic:
\be
K_C(N) \sim \big(\ln N\big)^{-1}, \quad \gamma=3 \;,
\label{K*N-log}
\ee
which has been observed in Monte Carlo simulations for the Ising model\cite{stauffer}.

To summarize we have studied the properties of first- and second-order phase transitions
of ferromagnetic lattice models on scale-free networks. Using the Weiss 
MF approximation we have derived a general SC equation that has been analyzed
by standard methods. Three regions of the phase diagram are identified as
a function of the degree exponent $\gamma$. First-order transitions
can only take place in the first regime, where $\gamma>\gamma^u$, and where the
effect of connectivity of the network is irrelevant. In the second regime,
for $3<\gamma \le \gamma^u$ the phase transition is always continuous
and takes place at a finite temperature. The magnetization critical exponent,
however, is nonconventional, $\gamma$ dependent. In the third regime, for
$\gamma \le 3$, the system is in its ordered phase at any temperature and
a singularity can develop only for diverging temperature (or vanishing
coupling).

As far as the properties of the critical singularities are concerned the MF
method presumably gives exact results.
The location of the transition point is not necessarily exact. Indeed, our results coincide
with others, obtained by different methods on
specific problems\cite{percolation,ising}. 

After this work has been completed we became aware of a preprint by
Goltsev, Dorogovtsev, and Mendes\cite{GDM02}, in which some results
about the $q$-state Potts model on the SF networks have been announced.
\vglue3mm
We are indebted to Jae Dong Noh for stimulating discussions.
The work by F.I. has been supported by the Hungarian National
Research Fund under  grants Nos. OTKA TO34183, TO37323,
MO28418, and M36803, by the Ministry of Education under grant No FKFP 87/2001,
by the EC Centre of Excellence (Grant No. ICA1-CT-2000-70029), and by the CNRS.
The Laboratoire de Physique des Mat\'eriaux
is Unit\'e Mixte de Recherche No.~7556.

\refrule

\end{multicols}

\begin{references}
\bibitem{note}
  This approximation is justified in special problems where
  the critical singularities of our method coincide with those
  obtained by other methods\cite{percolation,ising}.
\end{references}
\end{document}